\documentclass[epj]{svjour}

\usepackage{graphics}
\usepackage{graphicx}
\usepackage{stfloats}
\usepackage{flushend}
\usepackage{url}
\usepackage{ulem}
\usepackage{color}
\begin{document}
\title{
Particle productions and anisotropic flows from the AMPT model for Cu+Au collisions at $\sqrt{s_{NN}}=200$ GeV}
\author{Yuncun He\inst{1}\fnmsep\thanks{\email{hyuncun@foxmail.com}} \and Zi-Wei Lin\inst{2}\fnmsep\thanks{\email{linz@ecu.edu}}
}                     
%
\institute{Faculty of Physics and Electronic Technology, Hubei University, Wuhan 430062, China \and Department of Physics, East Carolina University, Greenville, NC 27858, USA}
\date{Received: date / Revised version: date}
%
\abstract{
We use the string melting version of a multi-phase transport (AMPT) model to study Cu+Au collisions at $\sqrt{s_{NN}}=200$ GeV. The rapidity distributions of identified hadrons show asymmetric dependences on rapidity. In addition, elliptic and triangular flows at mid-rapidity from the AMPT model for pions, kaons, and protons agree reasonably with the experimental data up to $p_{T}\sim1$ GeV$/c$. We then investigate the forward/backward asymmetry of $v_2$ and $v_3$. We find that these anisotropic flows are larger on the Au-going side than the Cu-going side, while the asymmetry tends to go away in very peripheral collisions.
We also make predictions on transverse momentum spectra of identified hadrons 
and longitudinal decorrelations of charged particles, where the average decorrelation of elliptic flow in asymmetric Cu+Au collisions is found to be stronger than that in Au+Au collisions.
} 
\authorrunning{He and Lin}
\titlerunning{Cu+Au collisions at $\sqrt{s_{NN}}=200$ GeV from the AMPT model}
\maketitle
\section{Introduction}
\label{intro}
Extensive studies indicate that heavy ion collisions at Relativistic Heavy Ion Collider (RHIC)  and the Large Hadron Collider (LHC) have created the deconfinement Quark-Gluon Plasma (QGP). Anisotropy flow is one of the well-known evidences to prove the creation of the hot and dense matter. It is the azimuthal anisotropy of final hadrons in the momentum space, which is converted from the initial spatial anisotropy  of the produced QGP. Hydrodynamical models~\cite{hyd-1,hyd-2,hyd-3} and transport models~\cite{trans-1,trans-2} have been successfully used to describe these signals. 
In addition, the longitudinal fluctuations affect the anisotropy flows, and the decorrelation of anisotropic flows at different pseudorapidities provides new insights into the initial condition along the longitudinal direction~\cite{Khachatryan:2015oea,Aaboud:2017tql,Pang:2014pxa,Wu:2018cpc}.
Large anisotropy flows have been observed even in small systems such as p+Pb collisions at LHC and d+Au collisions at RHIC~\cite{Aad:2012gla,Chatrchyan:2013nka,Adare:2014keg}. 
Although hydrodynamical models~\cite{Bozek:2013ska,Bozek:2012gr} and transport models~\cite{Bzdak:2014dia} can describe these phenomena, how exactly the anisotropy flows are generated is still an open question. 
Recent studies~\cite{Lin:2015ucn,He:2015hfa} have shown that the aniso-tropic parton escape may contribute to the flows more than the hydrodynamics evolution, especially for small systems. 

Asymmetric nuclei collisions such as Cu+Au provide another way to explore the initial state and parton interactions in heavy ion collisions. The collisions with unequal mass nuclei have asymmetric overlap geometry, energy density distribution and transverse length in the fireball. Investigation on Cu+Au collisions can help us get more information about the QGP and distinguish different models.

Event-by-event viscous hydrodynamics has been applied to predict the multiplicity, flow coefficients and femtoscopy radii in Cu+Au collisions at $\sqrt{s_{NN}}=200$ GeV~\cite{cuau-hyd}. 
The study on anisotropic flows in Cu+Au collisions with the previous AMPT model~\cite{cuau-ampt} showed that the directed flow is stronger and the elliptic flow is more sensitive to the parton cross section compared with Au+Au collisions. Recently, both PHENIX~\cite{cuau-ph} and STAR~\cite{cuau-star} Collaborations have published data on the azimuthal anisotropy at mid-rapidity in Cu+Au collisions at $\sqrt{s_{NN}}=200$ GeV based on the first asymmetric nuclei system runs in 2012. The PHENIX Collaboration also reported its measurements on the asymmetry of anisotropic flows~\cite{cuau-fb} and the nuclear modification factor~\cite{cuau-raa}. In this work, we use the string melting AMPT model with the new coalescence to calculate hadron productions in  Cu+Au collisions at $\sqrt{s_{NN}}=200$ GeV. In addition to comparing with the available experimental data on the elliptic flow and triangular flow at mid-rapidity, we also predict other observables such as the rapidity distribution, transverse momentum spectra, rapidity asymmetry of the elliptic flow and triangular flow, and longitudinal decorrelations.

\begin{figure*}[hpb]
\centering
\resizebox{1.05\textwidth}{!}{
 \includegraphics{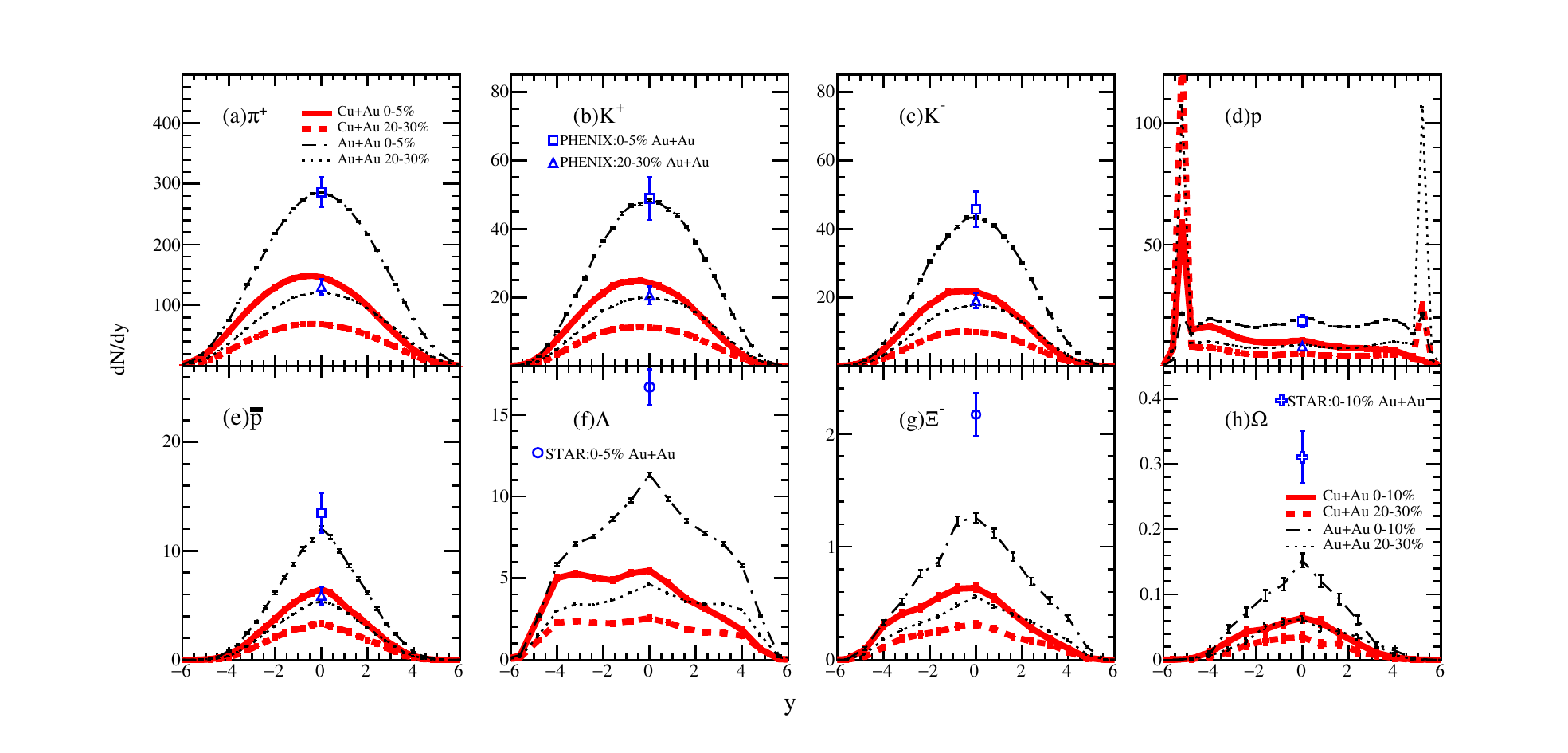}}
\caption{$dN/dy$ of identified hadrons in Cu+Au and Au+Au collisions at $\sqrt{s_{NN}}=200$ GeV from the AMPT model (curves) in comparison with the Au+Au data (symbols).}
\label{dndy}      
\end{figure*}

\section{The AMPT model with new coalescence}
\label{sec:1}
The AMPT model is a comprehensive event generator for heavy ion collisions~\cite{ampt}. The model incorporates the initial condition,  parton interactions, hadronization, and hadron interactions. The default AMPT model with the Lund string fragmentation as hadronization involves only minijet partons in the parton phase, and it can reproduce the yields and transverse momentum spectra of identified hadrons in heavy ion collisions at SPS and RHIC but fails to describe the elliptic flow at RHIC~\cite{Lin:2001zk}. 
On the other hand, the string melting AMPT model converts all excited strings into partons before the parton cascade and then describes the partonic matter to hadronic matter by the quark coalescence mechanism. Due to more energy and earlier interactions in the parton cascade, the string melting AMPT model can reasonably describe the anisotropic flows and long-range azimuthal correlations in A+A collisions and p+A collisions~\cite{Lin:2001zk,linprc,maprc}. However, the previous string melting AMPT model has problems in describing baryon productions; for example it produced more antiparticles than particles for multistrange baryons~\cite{maprc,prc-new}.

Recent development on the quark coalescence component~\cite{prc-new} 
has improved  baryon descriptions in the string melting AMPT model. 
The new coalescence has removed the previous constraint that forced the number of mesons, baryons and anti-baryons to be separately conserved in each event. As a result, each quark is free to coalesce to form a baryon or a meson, depending on the distance from the potential coalescence partners. 
As shown in that study~\cite{prc-new}, the new quark coalescence is not only more physical but also 
makes the coalescence to baryons more efficient.
In addition, the string melting AMPT model improved with this new coalescence can better describe the baryon yields, transverse momentum spectra and elliptic flows at low $p_{T}$ in both Au+Au collisions at $\sqrt{s_{NN}}=200$ GeV and Pb+Pb collisions at $\sqrt{s_{NN}}=2.76 $TeV~\cite{prc-new}. Therefore the AMPT model with the new quark coalescence is more reliable for the simulation of high-energy heavy  ion collisions, especially for baryons and antibaryons.

In the following, we will use the string melting AMPT model with the new quark coalescence to calculate Cu+Au collisions at $\sqrt{s_{NN}}=200$ GeV. We use the same main parameters as a recent study~\cite{prc-new}: the Lund string fragmentation parameters $a=0.55$  and $b=0.15$ GeV$^{-2}$~\cite{linprc}, the strong coupling constant $\alpha_s=$ 0.33, and a parton cross section of 1.5 mb. However, we terminate the hadron cascade at the global time of 300 fm/c because this study includes flows at large rapidities. 
Since hadron formation and interactions are time-dilated at large rapidities (roughly by the factor $\cosh y$~\cite{prc-new}) and hadronic rescatterings affect the flows of identified hadrons~\cite{mass-1,mass-2}, we need to use a large-enough global cutoff time to include hadronic scatterings at large rapidities. Note that centrality in this study is determined by the impact parameter distribution in minimum bias AMPT events.

\section{Particle yields and momentum spectra}
\label{sec:2}
Figure~\ref{dndy} shows the  $dN/dy$ yields of $\pi^{+}$, $K^{+}$, $K^{-}$, $p$, $\bar{p}$, $\Lambda$, $\Xi^{-}$ and $\Omega^{-}$ in Cu+Au collisions (thick curves) and Au+Au collisions (thin curves) at $\sqrt{s_{NN}}=200$ GeV. The curves represent AMPT results for most central 0-5\% collisions (but 0-10\% for $\Omega^{-}$) and semi-central 20-30\% collisions, while the symbols are experimental results for Au+Au collisions from PHENIX~\cite{Adler:2003cb} and STAR~\cite{star-auau1,star-auau2} Collaborations. 
Note that we simulate Cu+Au collisions of each centrality separately (e.g., $\sim $0.9 million events for 20-30\% centrality and $\sim $5 million events for the 40-60\% centrality), while the Au+Au results shown in this study are based on our previous calculations~\cite{prc-new}. 
We see that the AMPT model can well reproduce the mid-rapidity $dN/dy$ of $\pi^{+}$, $K^{+}$, $K^{-}$, $p$, and $\bar{p}$ in Au+Au collisions, although it underestimates the $dN/dy$ yields of strange baryons $\Lambda$, $\Xi^{-}$ and $\Omega^{-}$. The peak magnitude of $dN/dy$ in Cu+Au collisions is about half of that in Au+Au collisions, and the peaks are often shifted towards the backward rapidity (i.e. the Au-going side), consistent with the expectation of more participants in the Au-going side that results in a shift in the center-of-mass rapidity of the produced fireball~\cite{cuau-star}. However the shift is not obvious for $p$ and $\bar{p}$.
In addition, the rapidity distribution of pion, kaon, $\Lambda$, and $\Xi^{-}$ shifts more to the Au-going side in central collisions than in semi-central collisions~\cite{cuau-star}. Specifically, the mean rapidity 
in central collisions and semi-central collisions 
is about -0.43 and about -0.24 respectively for $\pi^{+}$, 
-0.39 and -0.23 respectively for $K^{+}$, -0.34 and -0.19 respectively for $K^{-}$, -0.13 and -0.08 respectively for $\bar{p}$, -0.78 and -0.42 respectively for $\Lambda$, -0.57 and -0.29 respectively for $\Xi^{-}$, -0.38 and -0.34 respectively for $\Omega^{-}$. For protons, the 
the meaning of the mean rapidity is less straightforward due to stopped protons from the incoming nuclei.
\begin{figure*}
\centering
\resizebox{1.05\textwidth}{!}{
\includegraphics{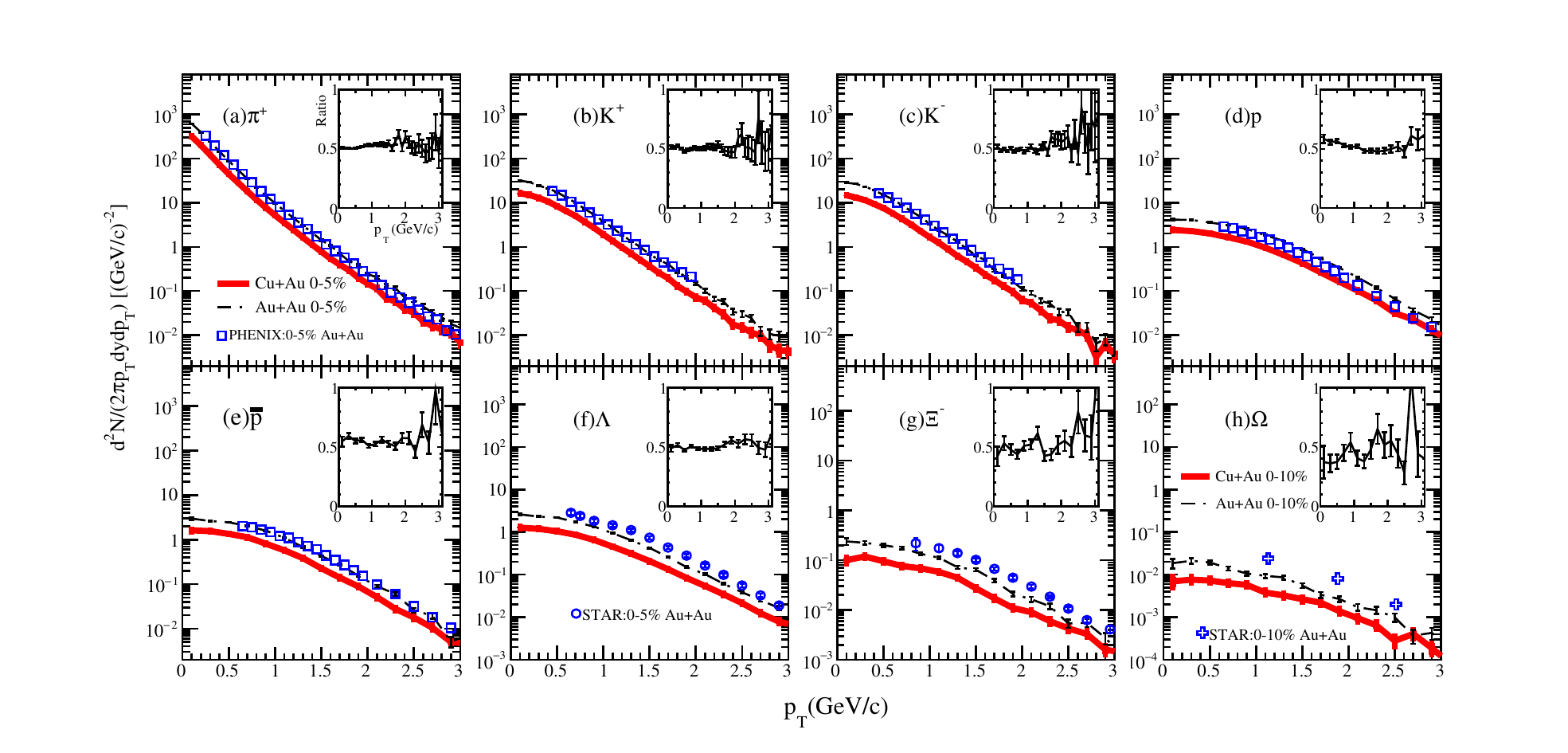}}
\caption{$p_{T}$ spectra at mid-rapidity for identified hadrons in central Cu+Au and Au+Au collisions at $\sqrt{s_{NN}}=200$ GeV from the AMPT model (curves) in comparison with the Au+Au data (symbols).}
\label{dndpt}      
\end{figure*}

We show in Fig.~\ref{dndpt} the $p_{T}$ spectra at mid-rapidity for $\pi^{+}$, $K^{+}$, $K^{-}$, $p$, $\bar{p}$, $\Lambda$, $\Xi^{-}$ and $\Omega^{-}$ in central Cu+Au collisions (thick curves) and central Au+Au collisions (thin curves) at $\sqrt{s_{NN}}=200$ GeV from the AMPT model in comparison with the Au+Au data. 
The centrality is 0-5\% for all the particles except that it is 0-10\% for $\Omega^{-}$. We also 
show the ratio of the spectrum in Cu+Au collisions to that in Au+Au collisions in the inset of each panel, and the ratio is $\sim 0.5$ for all the hadron species. The AMPT model can describe the Au+Au data of identified particles up to $p_{T} \sim 3$ GeV$/c$ of $\pi^{+}$, $K^{+}$, $K^{-}$, $p$, and $\bar{p}$.  
The shape of the $p_{T}$ spectrum at $p_{T}< 2$ GeV/c in central Cu+Au collisions for a given hadron species is about the same as that in central Au+Au collisions (except that protons are softer in Cu+Au collisions within $0<p_{T}<2$ GeV/c). We note that a study with hydrodynamics predicted that the transverse momentum spectra of  $\pi^{+}$, $K^{+}$, and $p$ are softer than that in Au+Au collisions~\cite{cuau-hyd}.

\begin{figure*}
\centering
\resizebox{0.9\textwidth}{!}{
\includegraphics[trim=0 0 0 20,clip]{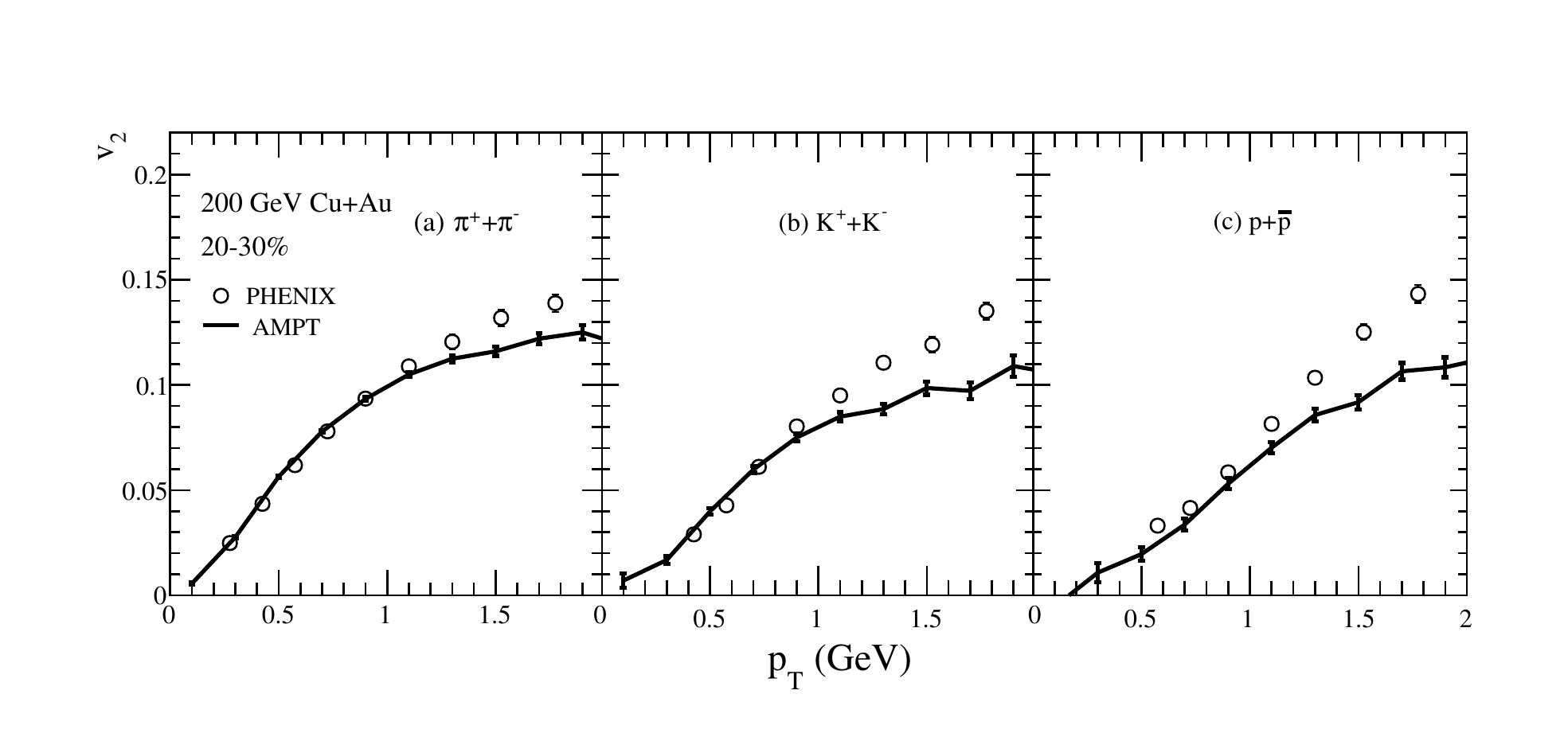}}
\vspace{-1em}
\caption{$v_2$ of $\pi$, $K$, and $p$ at mid-rapidity 
from the AMPT model in comparison with the PHENIX data
for 20-30\% Cu+Au collisions at $\sqrt{s_{NN}}=200$ GeV.}
\label{v2-pikp}       
\end{figure*}

\begin{figure*}
\centering
\resizebox{0.8\textwidth}{!}{
\includegraphics[trim=0 0 0 30,clip]{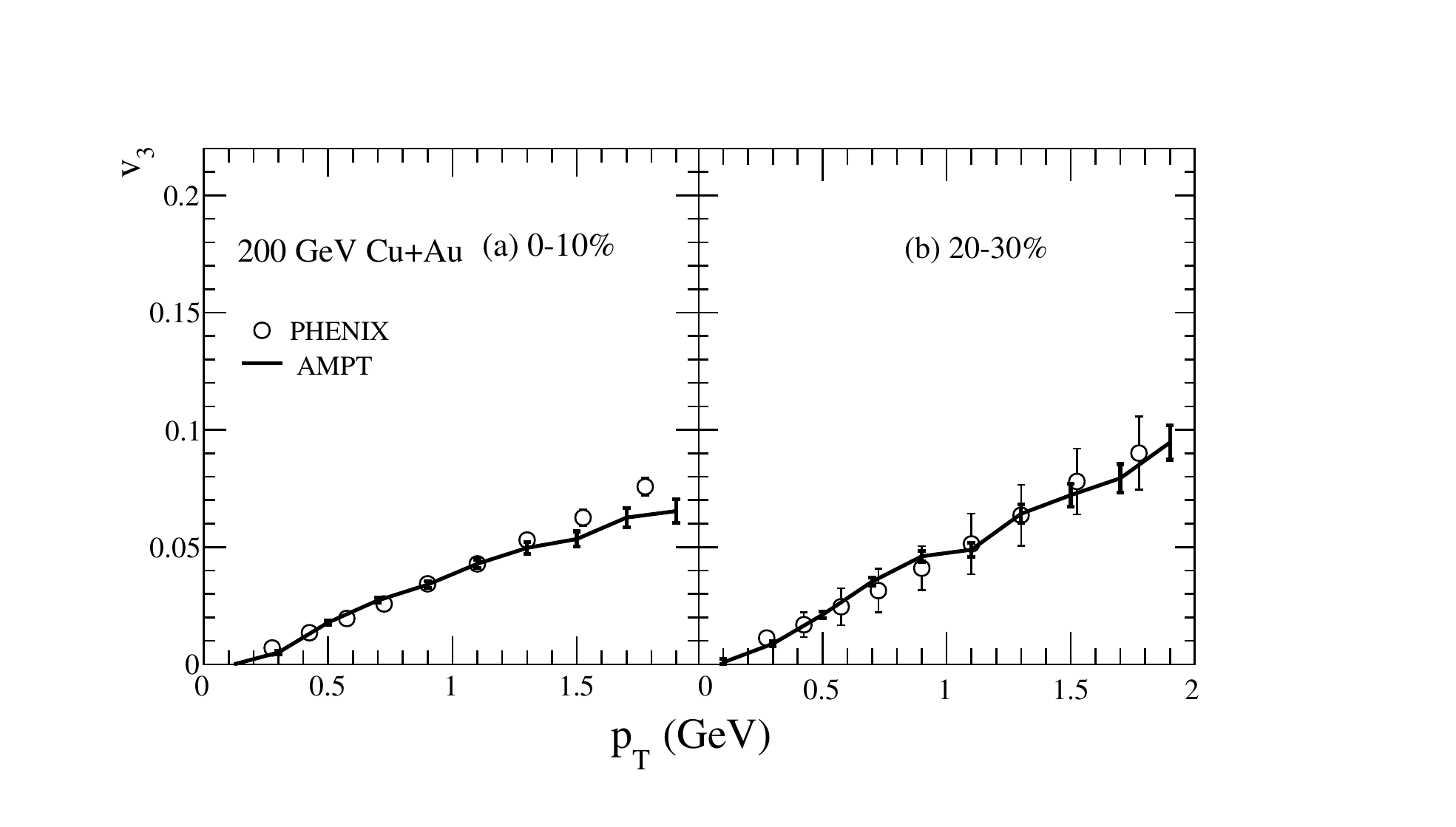}}
\vspace{-1em}
\caption{$v_3$ of charged hadrons at mid-rapidity 
from the AMPT model in comparison with the PHENIX data
for 0-10\% and 20-30\% Cu+Au collisions at $\sqrt{s_{NN}}=200$ GeV.}
\label{v3-h}       
\end{figure*}

\begin{figure*}
\centering
\resizebox{0.75\textwidth}{!}{
\includegraphics[trim=0 0 0 20,clip]{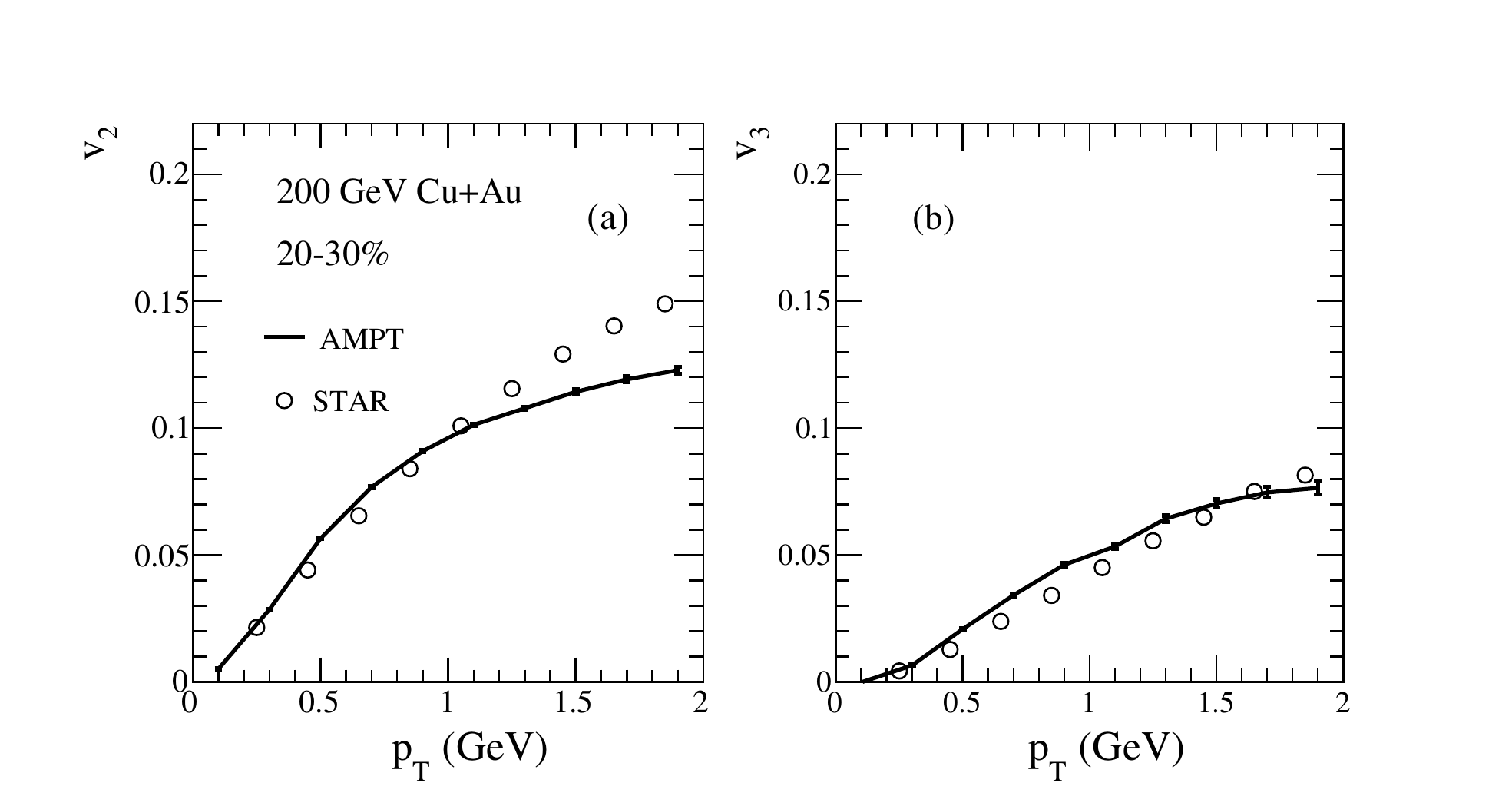}}
\caption{$v_2$ and $v_3$ of charged hadrons within $|\eta|<$ 1 in 
20-30\% Cu+Au collisions at $200$ GeV from the AMPT model 
in comparison with the STAR data.
}
\label{star}       
\end{figure*}

\section{Elliptic and triangular flows}
\label{sec:3}
The anisotropy coefficients $v_2$ and $v_3$ for the second and the third harmonic of the azimuthal particle distributions play an important role in the study of QGP properties~\cite{Heinz:2013th}. They are sensitive to the initial geometry and the properties of partonic and hadronic matter in heavy ion collisions~\cite{mass-1,mass-2,Alver:2010gr,Adare:2014bga}. In this study we calculate them as 
\begin{eqnarray}
v_{n}= \frac{\left<\rm cos[n(\phi-{\rm \Psi}_{n}^{\rm EP})]\right>}{\rm Res({\rm \Psi}_{n}^{\rm EP})},
\end{eqnarray}
where $\phi$ represents the azimuthal angle of a particle in momentum space, ${\rm \Psi}_{n}^{\rm EP}$ is the $n^{\rm th}$ harmonic event-plane, $\left <\cdots\right >$ 
denotes the averaging over particles in each event and then over all events,
and the denominator is the correction factor for the event-plane. In this study we calculate $v_n$ in Cu+Au collisions using the same method as the PHENIX Collaboration~\cite{cuau-ph}. For $v_n$ at mid-rapidity, we determine the event-plane ${\rm \Psi}_{n}^{\rm EP}$ with charged particles within 3 $<|\eta|<$ 3.9, and the resolution of the event-plane is calculated using the three-subevent method with particles in three pseudorapidity regions (-3.9 $<\eta<$-3, 3 $<\eta<$ 3.9, and $|\eta|<$ 0.35), where particles within $|\eta|<$ 0.35 are limited to $0.2<p_{T}<$2 GeV$/c$ to exclude jet contributions. For $v_n$ at large rapidities, we determine the event-plane ${\rm \Psi}_{n}^{\rm EP}$ with charged particles in the central bin ($|\eta|<$ 0.35) and use the three-subevent method for the correction factor of the event-plane~\cite{cuau-fb}.

Figure~\ref{v2-pikp} shows $v_2$ around mid-rapidity ($|\eta|<$ 0.35) as a function of $p_{T}$ from the AMPT model for 20-30\% Cu+Au collisions at $\sqrt{s_{NN}}=200$ GeV in comparison with PHENIX data~\cite{cuau-ph}. The left, middle, and right panels represent charged pions, charged kaons, and (anti)protons, respectively. We can see that AMPT can reproduce these identified hadrons' $v_2$ at low $p_{T}$. 
In Fig.~\ref{v3-h} we compare charged hadrons $v_3$
around mid-rapidity ($|\eta|<$ 0.35) with the PHENIX data 
for 0-10\% and 20-30\% Cu+Au collisions at $200$ GeV, where the AMPT results agree well with the data. The results here are consistent with an earlier calculation~\cite{cuau-ph} where the AMPT model with the old quark coalescence was used together with a modified Glauber model and a parton cross section of 3.0 mb.
As shown in Fig.~\ref{star}, $v_2$ and $v_3$ from the AMPT model with the new quark coalescence are also reasonably consistent with the mid-rapidity $|\eta|<$ 1 STAR data~\cite{cuau-star} at low $p_{T}$ for 20-30\% Cu+Au collisions at $200$ GeV. Here we use particles in three subevents ($-1<\eta<-0.4$, $|\eta|<0.2$, and $0.4<\eta<1$) 
with an $\eta$-gap of 0.4 and within $p_{T}<2$ GeV$/c$~\cite{cuau-star} to determine the event-plane and the event-plane resolution for comparison with the STAR data. 
Our results describe the STAR data better than the previous AMPT simulation with the old quark coalescence~\cite{cuau-star} which overestimated the flows at low $p_{T}$. Thus they provide a good baseline for us to further study the asymmetry of anisotropic flows in rapidity in Cu+Au collisions.

\begin{figure*}
\centering
\resizebox{0.9\textwidth}{!}{
\includegraphics[trim=0 0 0 20,clip]{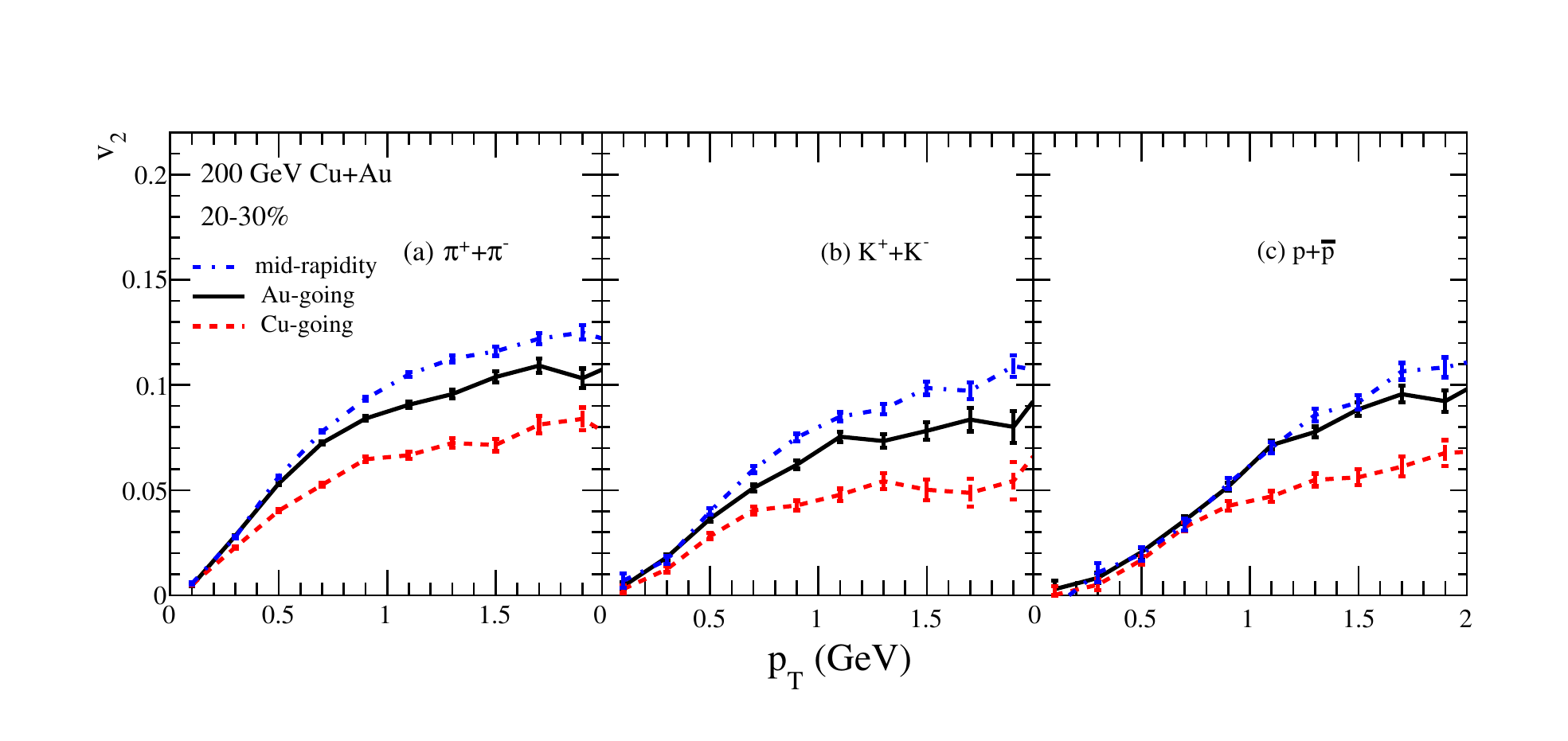}}
\vspace{-1em}
 \caption{$v_2$ of $\pi$, $K$, $p$ at mid-rapidity, backward and forward pseudorapidities for 20-30$\%$ Cu+Au collisions from the AMPT model.}
\label{v2-y}      
\end{figure*}

\begin{figure*}
\centering
\resizebox{0.75\textwidth}{!}{
\includegraphics[trim=0 0 0 20,clip]{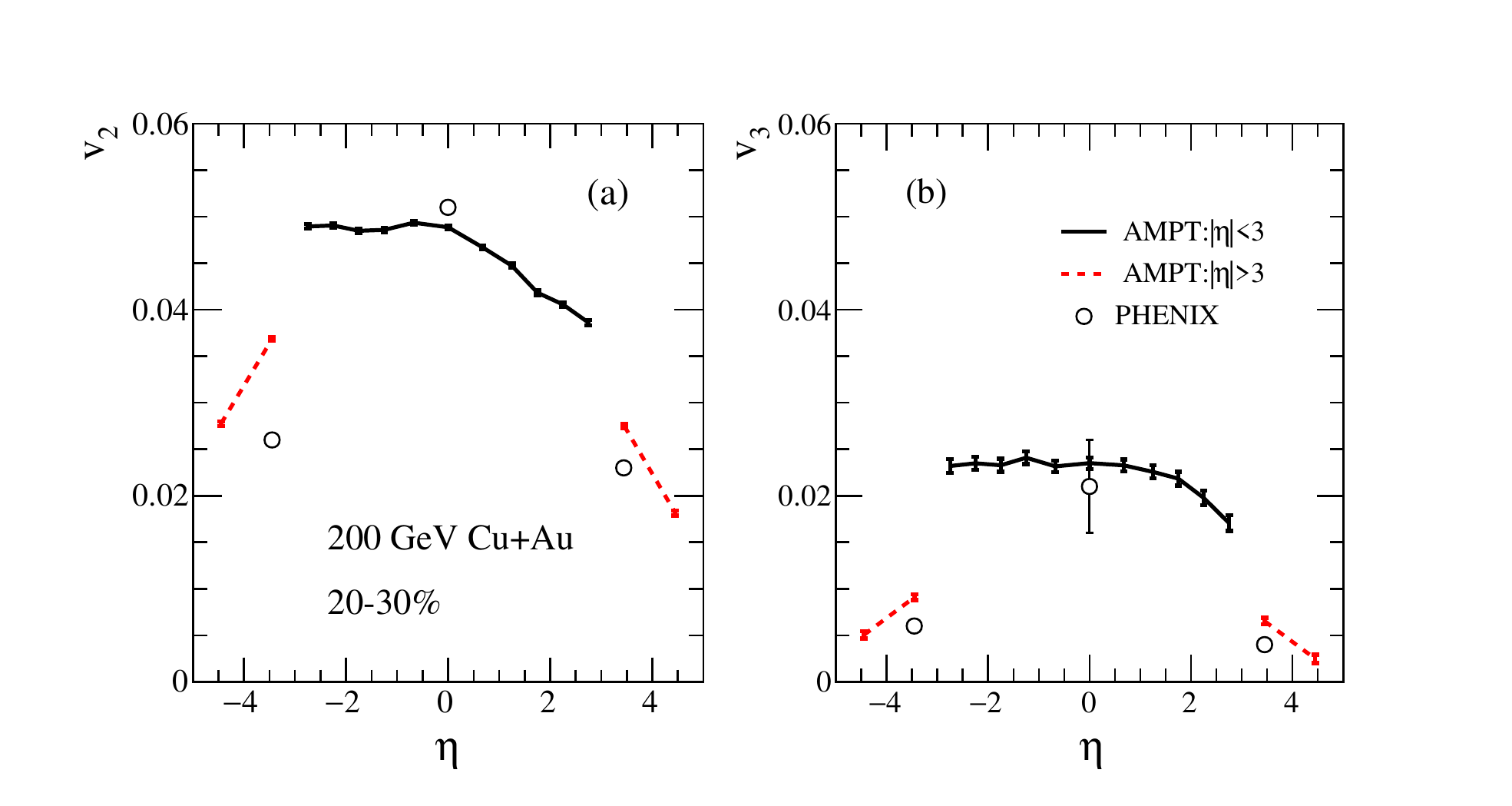}}
\caption{$v_2$ and $v_3$ of charged hadrons within $0<p_{T}<3$ GeV/c at different pseudorapidities from the AMPT model (curves) in comparison with the PHENIX data (circles) for 20-30$\%$ Cu+Au collisions.}
\label{v2-eta}       
\end{figure*}

A significant feature of Cu+Au collisions is the rapidity asymmetry as shown in Fig.~\ref{dndy}. In  Fig.~\ref{v2-y} we compare the $v_2$ of charged pions, charged kaons, and (anti)protons at large pseudorapidities 3 $<|\eta|<$ 3.9 and mid-rapidity in 20-30\% Cu+Au collisions. The solid curves represent the $v_2$ at backward pseudorapidities (i.e. the Au-going side), the dashed curves represent the $v_2$ at forward pseudorapidities (i.e. the Cu-going side), and the dot-dashed curves are results at mid-rapidity. We can see that the $v_2$ of these particles at mid-rapidity is usually the largest, while $v_2$ at the large backward pseudorapidity is stronger than that at the large forward pseudorapidity. This forward/backward asymmetry of $v_2$ in Cu+Au collisions should be related to asymmetry of the initial geometry between the Cu-going side and the Au-going side. This asymmetry is qualitatively similar to that in asymmetric small systems such as p+Pb collisions, where the anisotropic flows are larger in the Pb-going side than that in the proton-going side~\cite{ppb-fb}.

\begin{figure*}
\centering
\resizebox{\textwidth}{!}{
\includegraphics[trim=0 0 0 20,clip]{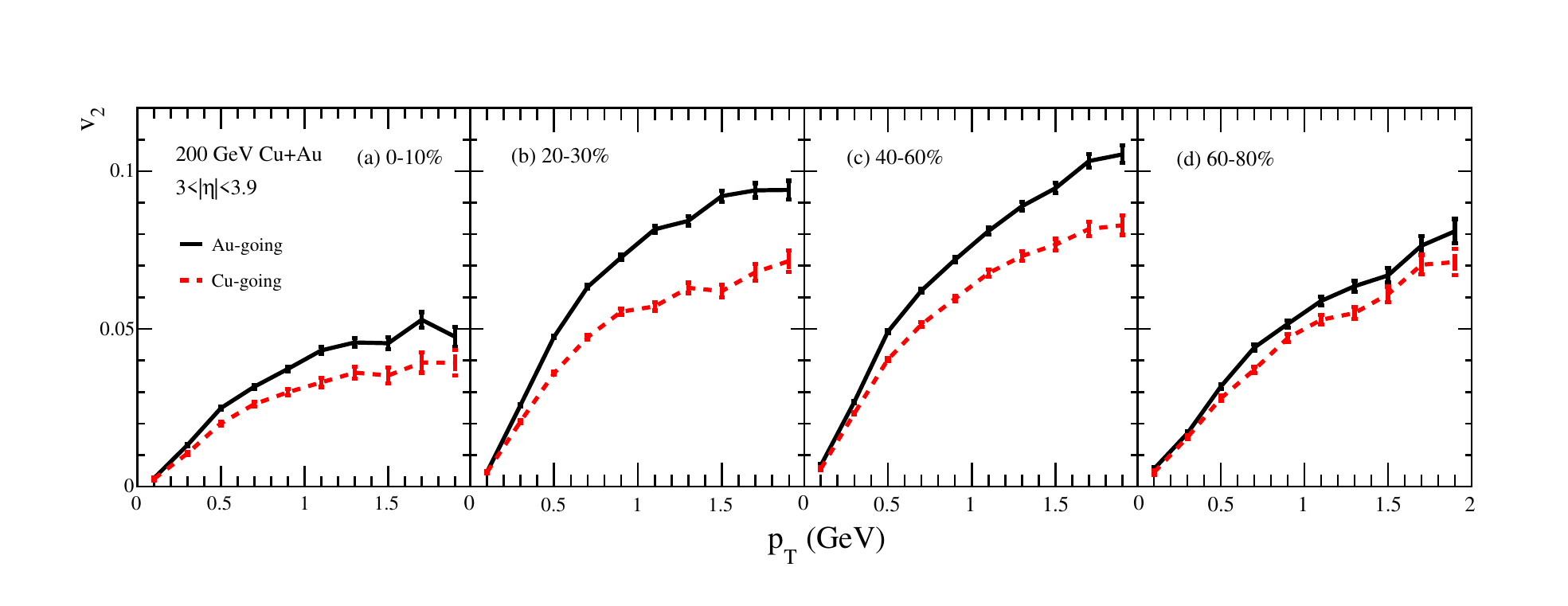}}
\vspace{-2em}
\caption{$v_2$ of charged particles at large backward and forward pseudorapidities in Cu+Au collisions at different centralities.}
\label{v2-4centrality}       
\end{figure*}

\begin{figure*}
\centering
\resizebox{\textwidth}{!}{
\includegraphics[trim=0 0 0 20,clip]{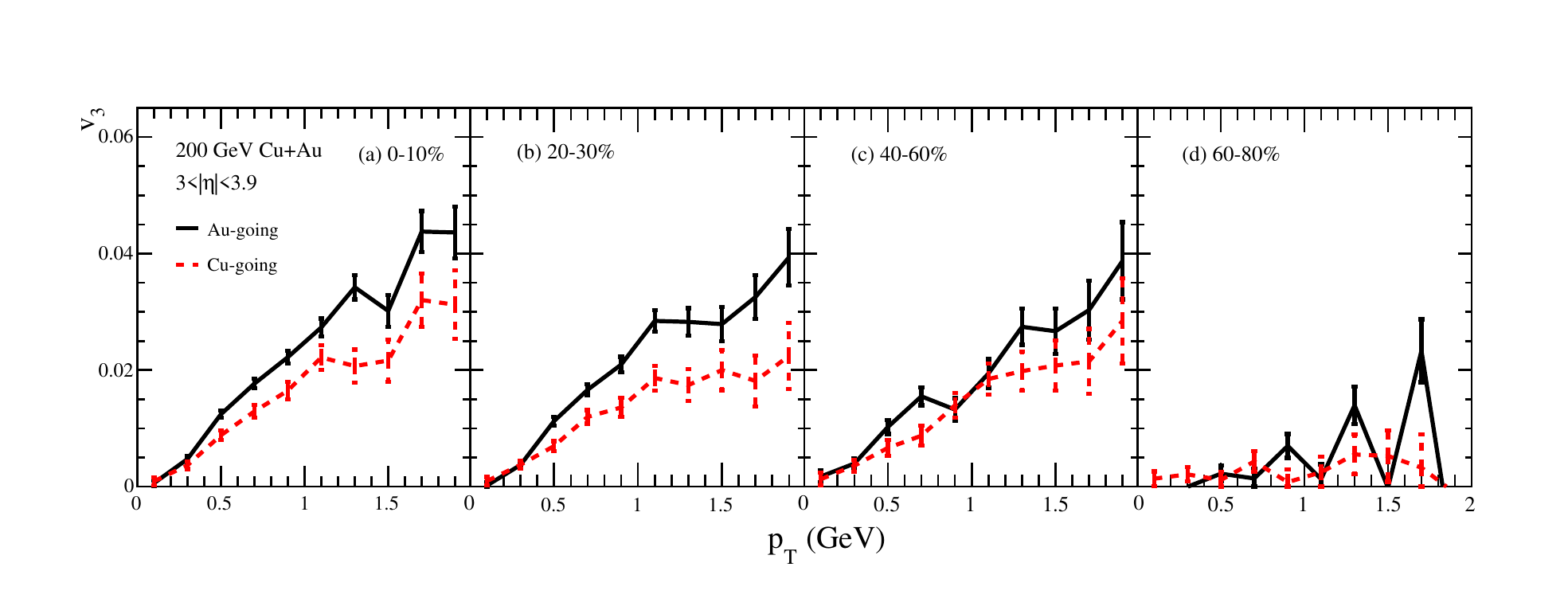}}
\vspace{-2em}
\caption{Same as Fig.~\ref{v2-4centrality} but for $v_3$.}
\label{v3-4centrality}       
\end{figure*}

\begin{figure*}
\centering
\resizebox{0.5\textwidth}{!}{
\includegraphics[trim=0 0 0 20,clip]{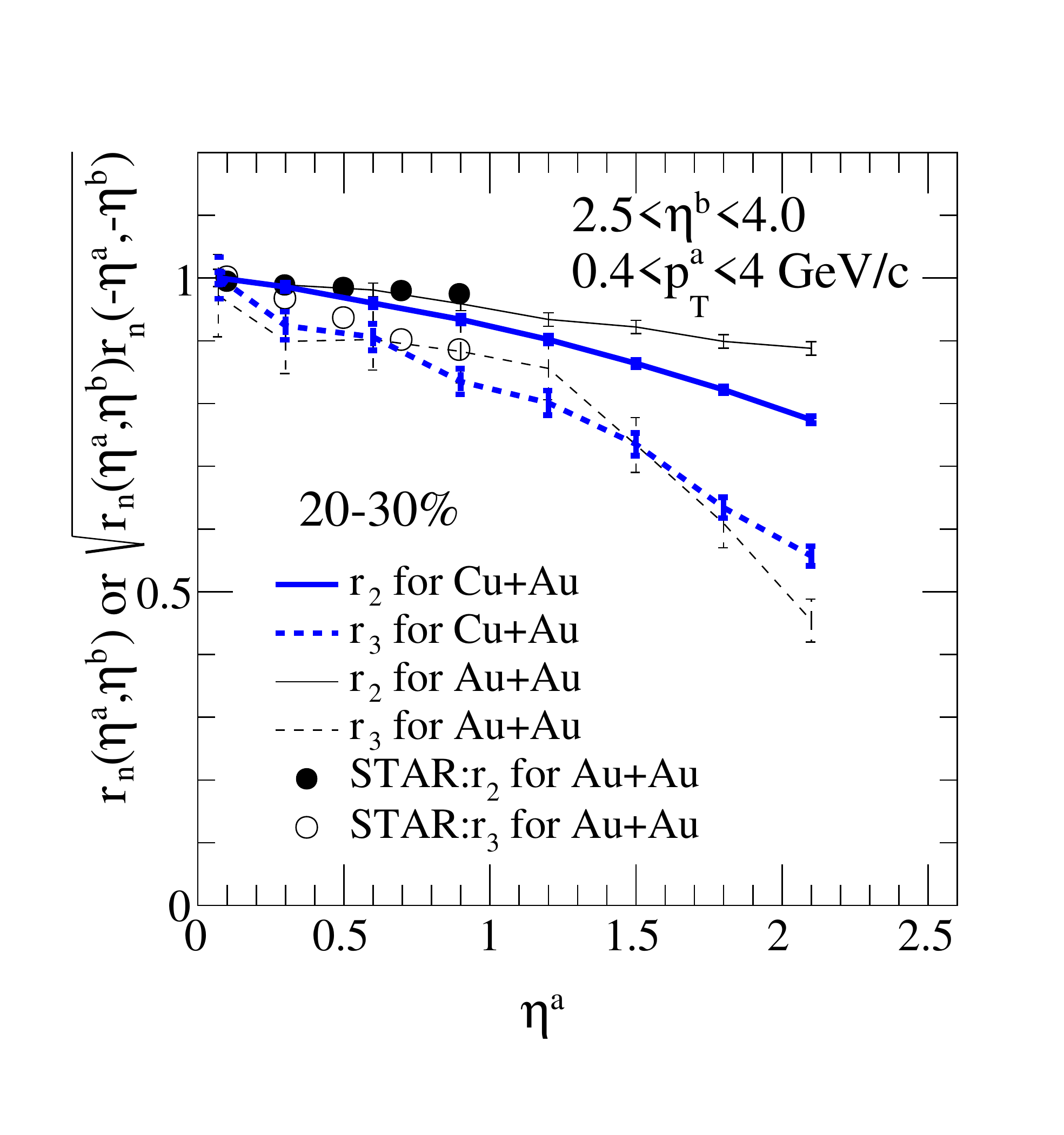}}
\vspace{-1em}
\caption{Factorization ratios $r_n(\eta^{a},\eta^{b})~(n=2,3)$ for 20-30$\%$ Au+Au collisions in comparison with the STAR data and the average factorization ratios $\sqrt{r_n(\eta^{a},\eta^{b})r_n(-\eta^{a},-\eta^{b})}~(n=2,3)$ for  20-30$\%$ Cu+Au collisions.}
\label{rn}       
\end{figure*}

The pseudorapidity dependences of $v_2$ and $v_3$ of charged hadrons within $0<p_{T}<3$ GeV/c are shown in Fig.~\ref{v2-eta} for 20-30\% Cu+Au collisions, where we have calculated the flows in different  pseudorapidity regions: 
$\eta \in$ (-5, -3.9), (-3.9, -3), (-3, -2.5), (-2.5, -2), (-2, -1.5), (-1.5, -1), (-1, -0.35), (-0.35, 0.35), (0.35, 1), (1, 1.5), (1.5, 2), (2, 2.5), (2.5, 3), (3, 3.9), and (3.9, 5). We use particles  within 3 $<|\eta|<$ 3.9 to determine the event-plane for the regions $|\eta|<$ 3 but use particles within $|\eta|<0.35$ for the $|\eta|>$ 3 regions. Solid curves in Fig.~\ref{v2-eta} 
are results for the regions $|\eta|<$ 3 while dashed curves are for the $|\eta|>$ 3 regions; the two sets of curves do not connect smoothly because of the above different event-plane methods and longitudinal decorrelation. They are compared with the PHENIX preliminary data in $|\eta|<$ 0.35 and  3 $<|\eta|<$ 3.9~\cite{qm2017,qm2017-2}.
The magnitudes of $v_2$ and $v_3$ at large pseudorapidities are higher than the PHENIX data, while the forward/backward asymmetry is roughly consistent with the PHENIX data. 
We also see that the flows in the backward pseudorapidity are larger than that in the forward pseudorapidity, 
and the peaks of the flows are in the backward pseudorapidity. 
This is consistent with the AMPT results in an earlier study~\cite{qm2017-3}, although there the peaks of the $v_n(\eta)$ curves usually appear closer to $\eta=0$ and the magnitudes of $v_n$ at large pseudorapidities are closer to the data.

\begin{table}
\centering
\caption{The ratio of average flow in backward pseudorapidity (-3.9 $<\eta<$ -3) over that in forward  pseudorapidity (3 $<\eta<$ 3.9) in Cu+Au collisions at different centralities.}
\label{tab1}       
\begin{tabular}{lll}
\hline\noalign{\smallskip}
Centrality & $\langle v_2(B)\rangle/\langle v_2(F)\rangle$ & $\langle v_3(B)\rangle/\langle v_3(F)\rangle$  \\
\noalign{\smallskip}\hline\noalign{\smallskip}
0-10\% & 1.30 $\pm$ 0.03 & 1.42 $\pm$ 0.08 \\
20-30\%& 1.35 $\pm$ 0.02 & 1.43 $\pm$ 0.10 \\
40-60\% &1.22 $\pm$ 0.02 & 1.34 $\pm$ 0.15 \\
60-80\% & 1.14 $\pm$ 0.03  &  0.71  $\pm$ 0.36 \\
\noalign{\smallskip}\hline
\end{tabular}
\end{table}

\begin{table}
\centering
\caption{Charged particle multiplicity density $dN_{ch}/d\eta$ in backward pseudorapidity, forward pseudorapidity, and their ratio for Cu+Au collisions at different centralities.}
\label{tab2}       
\begin{tabular}{lllc}
\hline\noalign{\smallskip}
Centrality & -3.9 $<\eta<$ -3 &  3 $<\eta<$ 3.9 & Ratio\\
\noalign{\smallskip}\hline\noalign{\smallskip}
0-10\% & 207.8 $\pm$ 0.5 & 124.4 $\pm$ 0.2  & 1.67 $\pm$ 0.01  \\
20-30\%& 103.3 $\pm$ 0.3 & 76.7 $\pm$ 0.2  & 1.35 $\pm$ 0.01  \\
40-60\% &36.7 $\pm$ 0.2 & 31.1 $\pm$ 0.2     &  1.18 $\pm$ 0.01  \\
60-80\% & 12.2 $\pm$ 0.1  &  10.0  $\pm$ 0.1  &  1.22 $\pm$  0.02 \\
\noalign{\smallskip}\hline
\end{tabular}
\end{table}

We show in Fig.~\ref{v2-4centrality} the AMPT results of charged particles' $v_2$ within 3 $<\eta<$ 3.9 and -3.9 $<\eta<$ -3 in Cu+Au collisions at different centralities. The difference of $v_2$ between backward and forward pseudorapidities first increases from central (0-10\%) collisions to semi-central (20-30\%) collisions and then decreases in more peripheral collisions. 
Fig.~\ref{v3-4centrality} shows similar features for the $v_3$ of charged particles; however, the statistical error bars are large. 
To better observe the forward/backward asymmetry of the flows in Cu+Au collisions, we present in Table~\ref{tab1} the ratio of the average flow magnitude in the backward pseudorapidity (-3.9 $<\eta<$ -3) over that in the forward pseudorapidity (3 $<\eta<$ 3.9) from the AMPT model. Note that the averaged flow is calculated by integrating $v_n(p_{T})$ with the $p_{T}$ spectrum as the weight. Table~\ref{tab1} shows that the averaged $v_2$ ratio has a clear dependence on the centrality, where the ratio is the highest for central (0-10\%) or semi-central (20-30\%) Cu+Au collisions but much lower at peripheral (60-80\%) collisions. 
The averaged $v_3$ ratios show similar dependence on the centrality, although the statistical errors are big.

To better understand the forward/backward asymmetry of flows in Cu+Au collisions, we list in Table~\ref{tab2} the charged  particle multiplicity density ($dN_{ch}/d\eta$) values within -3.9 $<\eta<$ -3, 3 $<\eta<$ 3.9 and their ratio at different centralities. The multiplicity density in the backward pseudorapidity is significantly larger than that in the forward pseudorapidity, thus we can expect the initial spatial anisotropy in the backward pseudorapidity to be better converted to  anisotropic flows in the momentum space. 
On the other hand, the centrality dependence of the backward/forward $dN_{ch}/d\eta$ ratios is not the same as that of the  $\langle v_n(B)\rangle/\langle v_n(F)\rangle$ ratios, because the initial spatial 
anisotropy can also have forward/backward asymmetry that consequently affects the asymmetry of flows. 

The forward/backward asymmetry of flows in Cu+Au collisions is also related to longitudinal  correlations and fluctuations. The longitudinal decorrelation has been measured at both LHC~\cite{Khachatryan:2015oea,Aaboud:2017tql} and RHIC~\cite{Nie:2019bgd}. The CMS Collaboration proposed the factorization ratio $r_{n}(\eta^{a},\eta^{b})$ to investigate the decorrelations~\cite{Khachatryan:2015oea}. For particles within $\eta^{a}$ and reference particles within $\eta^{b}$ , it is defined as
\begin{eqnarray}
r_{n}(\eta^{a},\eta^{b})= \frac{V_{n\Delta}(-\eta^{a},\eta^{b})}{V_{n\Delta}(\eta^{a},\eta^{b})},
\end{eqnarray}
where $V_{n\Delta}(\eta^{a},\eta^{b})=\left<\left< \rm cos(n\Delta\phi)\right>\right>$ and $\left<\left<\right>\right>$ denotes the averaging over all particle pairs in each event and then over all events. For asymmetric collision systems, the average factorization ratio can be defined~\cite{Khachatryan:2015oea} as the square root of the product of  $r_{n}(\eta^{a},\eta^{b})$ and $r_{n}(-\eta^{a},-\eta^{b})$, because it is not affected by the forward/backward asymmetry of $v_{n}$ but is only sensitive to the $\eta$-dependent event-plane correlations.

These factorization ratios for $n=2$ and 3 for Au+Au collisions and Cu+Au collisions 
at $\sqrt{s_{NN}}=200$ GeV in the 20-30\% centrality class are shown in Fig.~\ref{rn}. The pseudorapidity range for  reference particles is 2.5$<\eta^{b}<$4.0 while the particles in $\eta^a$ are within 0.4$<p_{T}^a<$4 GeV/c according to the STAR measurements~\cite{Nie:2019bgd}. 
Solid curves are the AMPT results for $r_{2}$ while dashed curves are for $r_{3}$. 
We see that the factorization ratios $r_{2}$ and $r_{3}$ in Au+Au collisions both agree reasonably well with the  STAR data within $\eta^{a}<$1~\cite{Nie:2019bgd}. 
The decrease of the average $r_2$ with $\eta^{a}$ in Cu+Au collisions is clearly faster than that in Au+Au collisions, consistent with the fact that there are fewer wounded nucleons (or excited Lund strings) for particle productions in Cu+Au collisions. 
On the other hand, the decrease of the average $r_3$ with $\eta^{a}$ in Cu+Au collisions 
is similar to that in Au+Au collisions while we note that the error bars in the Au+Au results are sizable.

\section{Summary}
\label{sec:4}
Using the AMPT model with the improved quark coalescence, we have investigated Cu+Au collisions at $\sqrt{s_{NN}}=200$ GeV. We predict the $dN/dy$ yields of identified hadrons and find that the peak of the yield shifts towards the Au-going side as expected. We also predict the transverse momentum spectra of identified hadrons and the longitudinal factorization ratios of charged particles. In addition, the elliptic flow and triangular flow at mid-rapidity from the AMPT model agree reasonably well with the experimental data at low $p_{T}$. Regarding the forward/backward asymmetry of elliptic and triangular flows, our results show that the flow on the Au-going side is stronger than that on the Cu-going side. The asymmetry in both $v_2$ and $v_3$ are the biggest at the 20-30\% centrality and then decline with the increase of centrality. These results on asymmetric nuclear collisions could help us better understand the initial stage and the later dynamics in relativistic heavy ion collisions.

 \begin{acknowledgement}
 This research is  supported by  the NSFC of China under Grant No. 11547016.
\end{acknowledgement}

%
%

\end{document}